\documentclass[referee]{aa}
\usepackage{txfonts}
\usepackage{graphicx}

\begin{document}

\title{ Chemical composition of UV-bright star ZNG 4 in the globular cluster
M13 \footnote {Based on observations obtained with the Subaru 8.2m Telescope, 
which is operated by the National Astronomical Observatory of Japan.}}

\author{ 
        S. Ambika \inst{1}
        \and M.Parthasarathy \inst{1}
        \and W.Aoki \inst{2}
        \and T.Fujii \inst{2}, \inst{3}
        \and Y.Nakada \inst{4}, \inst{5}
	\and Y.Ita \inst{4}
        \and H.Izumiura \inst{6} 
}

\institute{ 
Indian Institute of Astrophysics, Koramangala, Bangalore - 560034, India
\and National Astronomical Observatory, Mitaka, Tokyo 181-8588, Japan
\and Faculty of Science, Kagoshima University, 1-21-35 Korimoto, Kagoshima,
890-0065, Japan
\and Institute of Astronomy, The University of Tokyo, Mitaka, Tokyo, 181-0015,
Japan
\and Kiso Observatory, Institute of Astronomy, University of Tokyo, Mitake,
Kiso, Nagano 397-0101, Japan
\and Okayama Astrophysical Observatory, National Astronomical Observatory,
Kamogata, Okayama 719-0232, Japan
}

\date{ Received 20 May 2003 / Accepted 18 November 2003 }

\abstract
{
We present a detailed model-atmosphere analysis of  ZNG 4, a 
UV-bright star 
in the globular cluster M13. From the analysis of a high resolution 
($ R \approx 45,000 $) spectrum of the object, we derive the 
atmospheric parameters to be $\rm T_{eff}  =  8500\pm 250$ K, 
log g  $=$  2.5 $\pm$ 0.5 and $\rm [Fe/H]= -1.5$. Except 
for magnesium, chromium and strontium, all other even Z elements are 
enhanced with titanium and calcium being 
overabundant by a factor of 0.8 dex. Sodium is enhanced by a factor of 0.2 
dex. The luminosity of ZNG 4 and its position in the color-magnitude 
diagram of the cluster indicate that it is a Supra Horizontal Branch (SHB)  
(post-HB) star. 
The underabundance of He and overabundances of Ca, Ti, Sc and Ba in the
photosphere of ZNG 4 indicate that diffusion and radiative 
levitation of elements may be in operation in M 13 post-HB stars even at 
$\rm T_{eff}$ of 8500K. 
Detailed and more accurate abundance analysis of post-HB stars in several 
globular clusters is needed to further understand their abundance anomalies.

}

\authorrunning{S.Ambika}
\titlerunning{Chemical composition of UV-bright star ZNG 4 in M13 }
\maketitle

\section{Introduction}

The term "UV-bright stars" was introduced by Zinn et al.(1972)
for stars in globular clusters that lie above the horizontal branch (HB) and are
bluer than red giants. The name resulted from the fact that,
in the U band, these stars were brighter than all other
cluster stars. Further investigations showed that this group
of stars consist of blue horizontal branch (BHB) stars, supra horizontal 
branch stars (SHB), post asymptotic giant branch stars (post-AGB), 
post-early AGB (P-EAGB) stars and AGB-manque stars. (de Boer 1985, 1987, 
Sweigart et al.1974, Brocato et al. 1990, Dorman et al. 1993 and Gonzalez \&
Wallerstein 1994)

To derive the chemical composition of UV-bright stars in globular 
clusters
and to understand their evolutionary stages, we started a program
to obtain high resolution spectra of these objects in selected globular
clusters with the High Dispersion Spectrograph (HDS, Noguchi et al. 2002) of 
the 8.2m Subaru
Telescope. We selected a few UV-bright stars in the
globular cluster M13 from the papers of Zinn et al.(1972) and Harris
et al.(1983) to derive their chemical composition. In this paper we
report the analysis of a high resolution spectrum of the UV-bright star
ZNG 4 (RA ($16^{h}41^{m}37\fs528$) and DEC ($+36\degr30\arcmin43.86\arcsec $  ) (2000) ) 
(Zinn et al. 1972) in M13 
as the first target of our 
program.

M13 (NGC 6205) is a nearby well-studied globular cluster with a 
distance modulus of \mbox{$\rm (m-M)_0$} = 14\fm42 and metallicity 
of $\rm [Fe/H] = -1.51$ (Kraft and Ivans 2003). 
The position of ZNG 4 in the color-magnitude diagram of M13 (Paltrinieri 
et al. 1998) is shown in Figure 1. Many of the globular
clusters show a prominent gap in the blue tail of the HB, which is 
presumed to be due to differential mass loss on the Red Giant Branch (RGB). 
In M13 it is observed at $\rm T_{eff} = 10000 K$ (Ferraro et al. 1997).
High resolution spectroscopic studies of M13 BHB stars lying on either
side of the gap were carried out by Peterson et al.(1983, 1995) and
Behr et al.(1999, 2000a). They found anomalous photospheric
abundances in BHB stars. These photospheric anomalies are most likely 
due to diffusion - the gravitational settling of helium and radiative 
levitation of the metal atoms in the stable atmosphere of hot stars. 
They found variations in the photospheric abundances and rotational 
velocities of BHB stars as a function of their effective temperatures. 

\begin{figure}
\centering
\includegraphics[width=8.5cm]{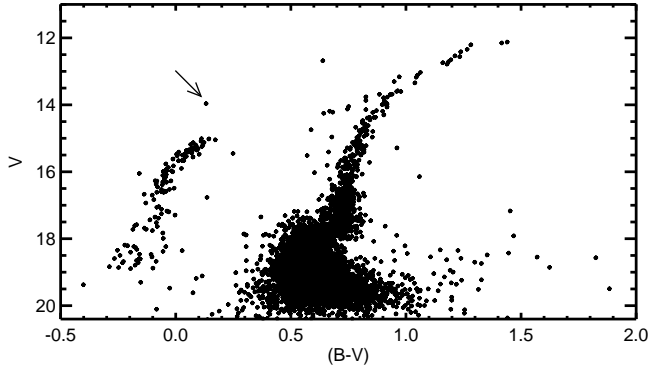}
\caption{
Color magnitude diagram (CMD) of globular cluster M13 obtained by
Paltrinieri et al.(1998). The arrow indicates the position of ZNG 4 in 
the CMD.}
\end{figure}

\section{Observations}

We have obtained a high resolution ($\rm \frac {\lambda} {\Delta \lambda}
\approx  45,000 $ ) spectrum of ZNG 4 on 15th (UT:14h45m) April 
2001 with the Subaru/HDS. The spectrum covering the
wavelength range 
4142 $\AA$- 6814 $\AA$ was obtained in an exposure time of 20 minutes.
There was no moon light problem during the observations and the sky 
background in the data was close to zero. We neglected the sky background 
in our data reduction.

The data was bias-subtracted, trimmed, flat-fielded to remove pixel
to pixel variations, converted to a one-dimensional spectrum, and normalized to
the continuum using standard CCD data reduction package (NOAO IRAF). The
spectrum has an average signal to noise ratio of 35. The reference 
spectrum of thorium-argon was used for the wavelength calibration. 

The various orders in our echelle spectrum of ZNG 4 have well defined
continuum and the normalization of the continuum was carried out using
the IRAF echelle spectra reduction programs. The continuum level in the
adjacent echelle orders to those containing the Balmer lines
was useful in defining the continuum in the Balmer line regions and the
profiles were normalized with a polynomial fit.

\section{Analysis}
The spectral lines were identified using Moore's atomic multiplet table (1945).
Equivalent widths of the absorption lines were measured using
the routines available in the SPLOT package of IRAF. The equivalent widths
were measured by Gaussian fitting to the observed profiles (and a multiple 
Gaussian fit to the blended lines such as the Mg II lines at 4481 $\AA$) 
and are given in Table 1.

\subsection{Radial velocity}

The radial velocity of ZNG 4 was derived from the wavelength shifts of many
absorption lines. The average heliocentric velocity is found to be 
$\rm V_r = -257.56 \pm 1.08$ $\rm km$ $\rm s^{-1}$ which 
is in agreement with the value derived by Zinn (1974) ($-$253 km $s^{-1}$).
It is also in agreement with the heliocentric velocities of M13 BHB stars
derived by Behr et al. (1999) and Moehler et al. (2003).

\subsection{Atmospheric parameters}

For the initial estimate of effective temperature, we looked for the
published CCD photometry of the star.
Recent CCD photometry of M13 was carried out by Rey et al. (2001).
However the ZNG 4 area of the cluster was not included in their observations
(Rey : private communication ). We used the published CCD photometry 
of ZNG 4 by Paltrinieri et al. (1998), who give, B$=$14.096 and V$=$13.964 . 
(B-V) = 0.132 and E(B-V) = 0.02 (Kraft and Ivans 2003) will yield
$\rm (B-V)_o =  0.112 $ which corresponds to $\rm T_{eff} = 8373 K$
(Flower 1996). However, the $ \rm (B-V)_0$ and $\rm T_{eff}$ calibration 
given by Flower (1996) is for Population I stars.

For our analysis, excitation potential and oscillator strengths of the lines 
were taken from the Vienna Atomic Line Database 
( http://www.astro.univie.ac.at/~vald/ ).
We employed the latest (2002) version of MOOG, an 
LTE stellar line analysis program (Sneden 1973) and Kurucz (1993) grid of 
ATLAS models. 
MOOG has been used successfully in the analysis of the spectra of warmer 
stars with $\rm T_{eff} = 7900$ K ( Preston and Sneden 2000).

We have also analyzed the spectra using the Kurucz WIDTH program (Kurucz CDROM
13, 1993) for verification. We used the line list obtained using
version 43 of the Synspec code of Hubeny and Lanz which is
distributed as part of their TLUSTY model atmosphere program.
( http://tlusty.gsfc.nasa.gov/Synspec43/synspec-line.html ) and also 
the information from the Kurucz linelist 
( http://kurucz.harvard.edu/linelists.html ).

The value of effective temperature was obtained by the method of excitation
balance, forcing the slope of abundances from Fe I lines versus excitation
potential to be zero. The surface gravity was then set by ionization\--
equilibrium, forcing abundances obtained from neutral (Fe I) and ionized 
(Fe II) species to be equal. The microturbulent velocity was estimated by
demanding that there should be no dependence of the Fe I abundance upon
equivalent widths of Fe I lines. 

\begin{figure*}
\centering
\includegraphics[angle=90,width=12.9cm]{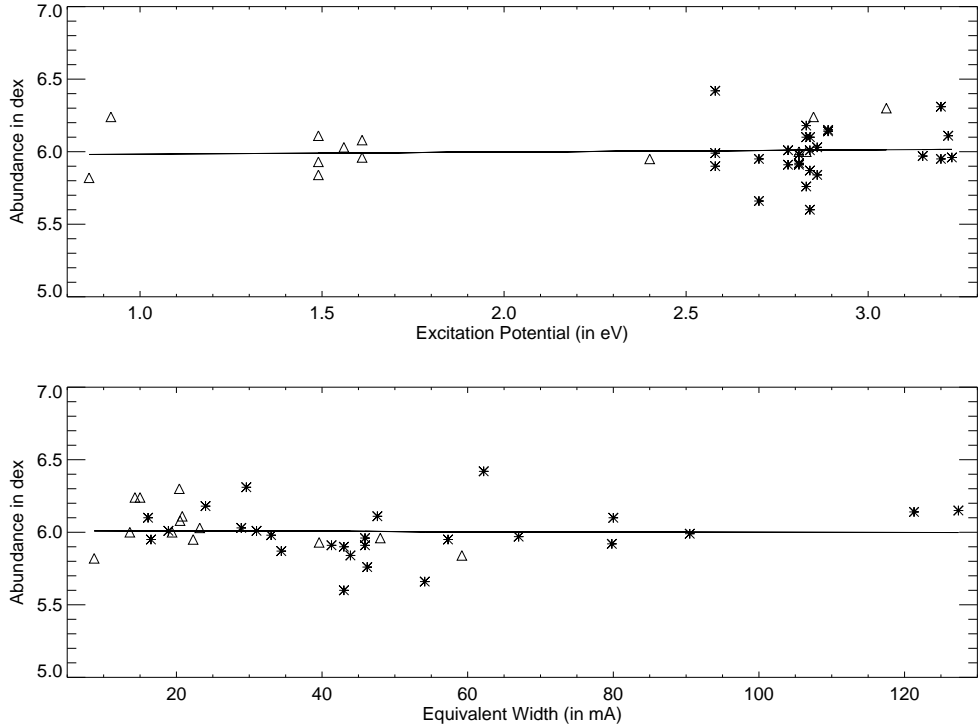}
\caption{
Top figure is the plot of the abundances from Fe lines versus 
excitation potential of the lines. Figure at the bottom is the plot of
the abundances from Fe lines versus their equivalent widths. Triangles
represent the Fe I lines and stars denote the Fe II lines.
}
\end{figure*}

The plots of the abundances versus excitation potentials and abundances 
versus equivalent widths in the case of Fe I and Fe II lines are shown 
in Figure 2. Such plots were made by varying the $\rm T_{eff}$, log g
and $\rm V_{t}$ in steps of 250 K, 0.5 and 0.5 km $\rm s^{-1}$ respectively
to estimate the uncertainties in these parameters.

From our analysis, we find that $\rm T_{eff} = 8500 K$, log g = 2.5 and
$\rm V_t=2.5 km s^{-1}$ fit the data best (Figure 2).  From the 
abovementioned method of analysis we find the uncertainties in $\rm T_{eff}$
to be 250 K, log g $=$ 0.5 dex and $\rm V_{t}$ $=$ 0.5 km $\rm s^{-1}$.
Uncertainties in derived abundances as a result of errors in the 
determination of the parameters and errors in the measurements
of equivalent widths are found to be of the order of 0.2 dex.

Using the derived atmospheric parameters and abundances, a 
synthetic spectrum was generated and plotted over the observed spectrum for 
verification. The observed and synthetic spectra were found to match well 
with the above mentioned atmospheric parameters and the final abundances 
are given in 
Tables 1 and 2. A region of the observed and synthetic spectrum is shown 
in Figure 3. The abundances derived using MOOG (Tables 1 and 2) are in good
agreement with the abundances derived using the WIDTH (Table 2).

\setcounter{table}{0}
\begin{table*} [ht]
\centering
\renewcommand{\thetable}{\arabic{table}}
\caption{Data for spectral lines measured in the spectrum of ZNG 4 in M13}
\begin{tabular}{lllllll} \hline

$\lambda_{lab}$ (in $\AA$)&LEP (eV)&log gf&EW ($m\AA$)&$\rm \log \epsilon $\\ \hline
He I\\
4471.47&20.96&-0.278&11.3&10.44\\
Na I\\
5889.95&0.00&0.117&63.9&5.05\\
5895.92&0.00&-0.184&40.6&5.04\\
Mg I\\
5167.32&2.71&-1.030&31.9&6.19\\ 
5172.68&2.71&-0.402&70.2&6.09\\
5183.60&2.72&-0.180&85.8&6.07\\
Mg II\\
4481.13&8.86&0.740&104.1&6.12\\ 
4481.32&8.86&0.590&80.0&5.92\\
Si II\\
5041.02&10.07&0.291&27.5&6.59\\ 
5055.98&10.07&0.593&25.5&6.24\\ 
6347.11&8.12&0.297&84.2&6.49\\ 
6371.37&8.12&-0.003&64.1&6.48\\  
Ca I\\
4226.73&0.00&0.265&76.7&5.21\\ 
4454.78&1.90&0.335&27.7&5.71\\
Ca II\\
5019.97&7.51&-0.501&18.6&5.77\\
Sc II\\
4246.82&0.32&0.242&54.5&1.58\\
4314.08&0.62&-0.096&47.8&2.03\\
4320.73&0.60&-0.252&49.7&2.21\\
4325.00&0.59&-0.442&35.9&2.18\\
4374.46&0.62&-0.418&39.8&2.23\\
4400.39&0.60&-0.536&32.3&2.21\\
4415.56&0.59&-0.668&31.5&2.32\\
Ti II\\
4161.53&1.08&-2.160&20.8&4.27\\ 
4163.65&2.59&-0.210&76.3&4.22\\ 
4171.91&2.60&-0.270&75.0&4.27\\  
4287.87&1.08&-1.820&38.8&4.26\\ 
4290.22&1.17&-0.930&91.1&4.16\\ 
4294.10&1.08&-0.880&95.2&4.12\\ 
4300.05&1.18&-0.490&129.9&4.43\\  
4301.91&1.16&-1.200&78.7&4.25\\ 
4307.86&1.17&-1.100&62.6&3.93\\ 
\hline
\end{tabular}
\end{table*}

\setcounter{table}{0}
\begin{table*} [ht]
\centering
\renewcommand{\thetable}{\arabic{table}}
\caption{continued}
\begin{tabular}{lllllll} \hline
$\lambda_{lab}$ (in $\AA$)&LEP (eV)&log gf&EW ($m\AA$)&$\rm \log \epsilon $\\ \hline
Ti II\\
4312.86&1.18&-1.090&77.5&4.13\\
4314.98&1.16&-1.120&69.8&4.04\\
4320.96&1.17&-1.900&43.4&4.47\\
4330.24&2.05&-1.800&21.6&4.57\\
4330.69&1.18&-2.060&24.3&4.30\\
4344.29&1.08&-1.930&19.8&3.99\\
4350.83&2.06&-1.810&12.6&4.32\\ 
4367.66&2.59&-0.870&34.9&4.28\\
4386.84&2.60&-0.940&27.2&4.21\\
4391.03&1.23&-2.240&19.2&4.38\\
4394.05&1.22&-1.770&37.7&4.28\\
4395.03&1.08&-0.510&142.4&4.62\\
4395.85&1.24&-1.970&31.7&4.39\\
4399.77&1.24&-1.220&78.3&4.31\\
4407.68&1.22&-2.430&10.4&4.27\\
4411.07&3.10&-0.670&23.0&4.18\\
4417.72&1.17&-1.230&76.0&4.23\\
4418.33&1.24&-1.990&25.0&4.27\\
4421.94&2.06&-1.580&14.6&4.16\\
4443.79&1.08&-0.700&118.3&4.33\\
4450.48&1.08&-1.510&56.7&4.20\\ 
4464.45&1.16&-1.810&39.8&4.31\\ 
4488.33&3.12&-0.510&40.0&4.36\\ 
4529.47&1.57&-1.650&27.1&4.21\\ 
4533.97&1.24&-0.540&126.3&4.42\\
4549.62&1.58&-0.220&102.1&3.89\\  
4563.76&1.22&-0.790&108.1&4.32\\ 
4571.97&1.57&-0.230&69.9&3.42\\ 
4589.96&1.24&-1.620&38.7&4.14\\ 
4763.88&1.22&-2.360&18.9&4.47\\
4779.98&2.05&-1.260&32.9&4.24\\  
4798.52&1.08&-2.670&12.9&4.49\\ 
4805.09&2.06&-0.960&52.1&4.24\\ 
4874.01&3.10&-0.900&18.0&4.27\\ 
4911.19&3.12&-0.650&28.4&4.28\\ 
5129.15&1.89&-1.300&33.4&4.17\\ 
5154.07&1.57&-1.780&26.9&4.29\\ 
5185.91&1.89&-1.370&26.2&4.10\\ 
5188.68&1.58&-1.050&68.5&4.20\\ 
\hline
\end{tabular}
\end{table*}

\setcounter{table}{0}
\begin{table*} [ht]
\centering
\renewcommand{\thetable}{\arabic{table}}
\caption{continued}
\begin{tabular}{lllllll} \hline
$\lambda_{lab}$ (in $\AA$)&LEP (eV)&log gf&EW ($m\AA$)&$\rm \log \epsilon $\\ \hline
Ti II\\
5226.54&1.57&-1.230&52.1&4.16\\
5336.77&1.58&-1.630&31.9&4.25\\
5381.02&1.57&-1.970&21.9&4.38\\
Cr II\\
4554.99&4.07&-1.282&14.8&4.40\\
4558.65&4.07&-0.449&44.5&4.18\\
4588.20&4.07&-0.627&34.6&4.20\\
4616.63&4.07&-1.361&15.0&4.48\\
4618.80&4.07&-0.840&26.1&4.25\\
4634.07&4.07&-0.990&19.1&4.23\\
4824.13&3.87&-0.970&20.6&4.11\\
5237.33&4.07&-1.160&13.0&4.21\\
Fe I\\
4143.87&1.56&-0.511&23.2&6.03\\
4199.10&3.05&0.155 &20.4&6.30\\
4202.03&1.49&-0.708&20.8&6.11\\
4260.47&2.40&0.109 &22.3&5.95\\
4271.76&1.49&-0.164&39.6&5.93\\
4325.76&1.61&0.006 &48.0&5.96\\
4383.55&1.49&0.200 &59.2&5.84\\
4415.12&1.61&-0.615&20.5&6.08\\
4891.49&2.85&-0.112&15.0&6.24\\
4920.50&2.83&0.068 &13.6&6.00\\
4957.60&2.81&0.233 &19.4&6.00\\
5269.54&0.86&-1.321& 8.7&5.82\\
5328.04&0.92&-1.466&14.3&6.24\\
Fe II\\
4173.46&2.58&-2.740&62.2&6.42\\
4178.86&2.58&-2.500&43.0&5.90\\
4233.17&2.58&-1.900&90.5&5.99\\
4296.57&2.70&-3.010&16.5&5.95\\
4351.77&2.70&-2.020&54.1&5.66\\
4385.39&2.78&-2.680&31.0&6.01\\
4416.83&2.78&-2.410&41.3&5.91\\
4489.18&2.83&-2.970&24.0&6.18\\
4491.40&2.86&-2.700&28.9&6.03\\
4508.29&2.86&-2.250&43.9&5.84\\
4515.34&2.84&-2.450&34.4&5.87\\
4520.22&2.81&-2.600&33.0&5.98\\
\hline
\end{tabular}
\end{table*}

\setcounter{table}{0}
\begin{table*} [ht]
\centering
\renewcommand{\thetable}{\arabic{table}}
\caption{Continued}
\begin{tabular}{lllllll} \hline
$\lambda_{lab}$ (in $ \AA$)&LEP (eV)&log gf&EW ($m \AA $)&$\rm \log \epsilon $\\ \hline
Fe II\\
4522.63&2.84&-2.030&43.0&5.60\\
4549.47&2.83&-2.020&80.0&6.10\\
4555.89&2.83&-2.160&46.2&5.76\\
4576.34&2.84&-2.920&18.9&6.01\\
4582.84&2.84&-3.090&16.1&6.10\\
4583.84&2.81&-1.860&79.8&5.92\\
4629.34&2.81&-2.330&45.9&5.91\\
4923.93&2.89&-1.320&121.3&6.14\\
5018.44&2.89&-1.220&127.4&6.15\\
5169.03&2.89&-1.303&134.4&6.37\\
5197.58&3.23&-2.100&45.9&5.96\\
5234.62&3.22&-2.230&47.6&6.11\\
5276.00&3.20&-1.940&57.3&5.95\\
5316.61&3.15&-1.850&67.0&5.97\\ 
5362.87&3.20&-2.739&29.6&6.31\\ 
6456.38&3.90&-2.100&45.8&6.46\\ 
Sr II\\
4215.52&0.00&-0.145&19.8&0.71\\ 
Y II\\
4177.53&0.41&-0.160&8.98&1.09\\
4374.94&0.41&0.160&11.10&0.86\\ 
Ba II\\
4554.03&0.00&0.170&17.03&0.95\\
4934.08&0.00&-0.150&8.042&0.88\\
\hline
\end{tabular}
\end{table*}

\begin{figure*}
\centering
\includegraphics[width=13.9cm]{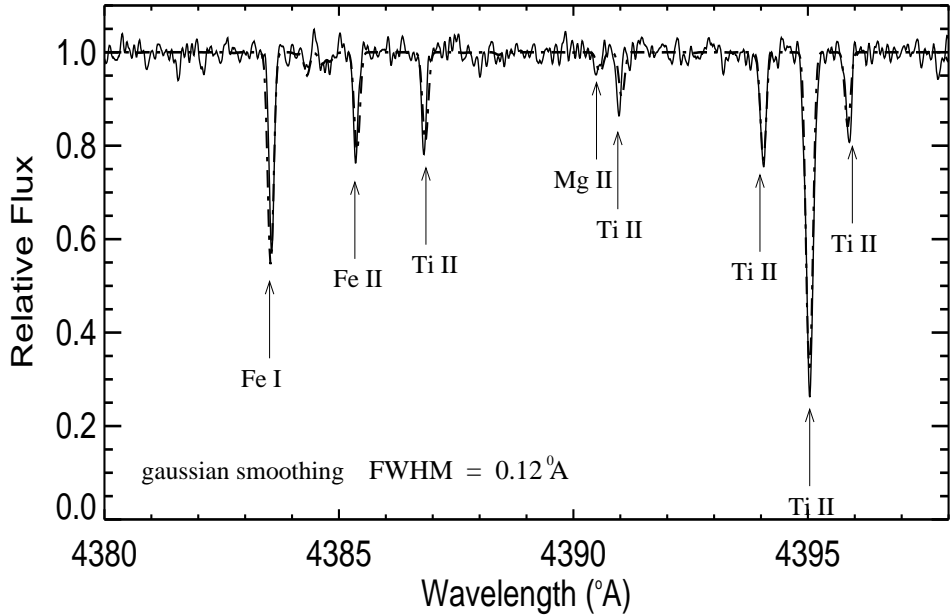}
\caption{Synthetic spectrum calculated with the atmospheric parameters 
($\rm T_{eff} = 8500$ K, log g = 2.5, $\rm V_t = 2.5$ km $s^{-1}$ ) and 
abundances ( Tables 1 and 2) is overplotted on the observed spectrum in the 
4380 $\AA$ - 4400 $\AA$ region.}
\end{figure*}

\subsection {Balmer Lines}

We tried to estimate the $\rm T_{eff}$ and log g from 
the analysis of Balmer lines in the spectrum of ZNG4 using
the Kurucz spectral atlas for Balmer lines (Kurucz CDROM 13, 1993). 

We could not get a satisfactory fit between the observed and theoretical 
Balmer line profiles with the atmospheric parameters $\rm T_{eff} = 8500$ K, 
log g $=$ 2.5, $\rm V_t = 2.5$ km $\rm s^{-1}$ and $\rm [Fe/H] = -1.5$ .
We also tried models that take into account the alpha element enhancements
and no convective overshooting 
(ANOVER models: HTTP://KURUCZ.harvard.edu/grids/gridm15ANOVER/ bm15ak2nover.dat). However, the profiles were found to be similar (Figure 4).
The best fit to $\rm H_{\beta }$ profile was obtained with the parameters
$\rm T_{eff} = 8750$ K, log g $=$ 2.0,  $V_t = 2.0$
km $\rm s^{-1}$ and $\rm [Fe/H] = - 1.5$ (Figure 4).
However, the excitation balance and ionization equilibrium for Mg and Fe lines 
could not be achieved with the above
parameters (see the last column in Table 2) and the abundances of Mg I, Mg II and Fe I, Fe II were found to differ
significantly (Table 2).
Therefore, we chose the model atmosphere determined from the
analysis of metal lines ( ie $\rm T_{eff} = 8500$ K, log g $=$ 2.5, 
$\rm V_t = 2.5$ km $\rm s^{-1}$ and $\rm [Fe/H] = - 1.5$ )
to represent the atmosphere of the star. 
%

The problem of fitting Balmer line profiles of HB stars has been mentioned 
by Grundahl et al. (1999) (and references therein).
For stars being more luminous than HB stars, mass loss and/or extended
atmosphere may influence the Balmer line profiles (Vink \& Cassisi, 2002).

\begin{figure*}
\centering
\includegraphics[width=8.5cm, angle=90]{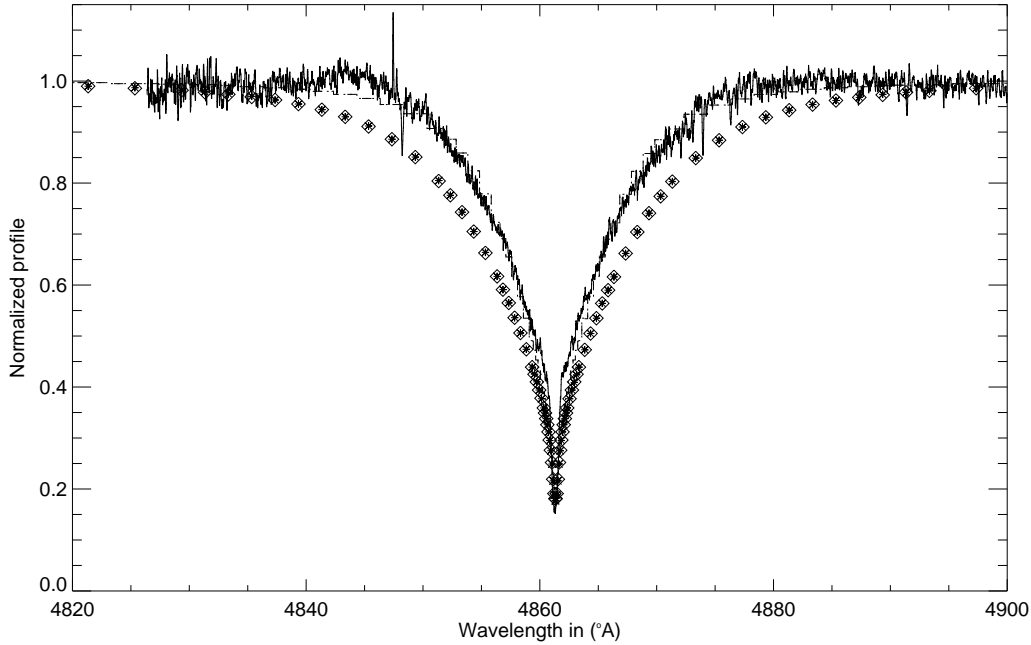}
\caption{Observed $\rm H_{\beta}$ profile compared with theoretical 
$\rm H_{\beta}$ profiles for 2 
different model atmospheric parameters. Dots and dashes represent the 
model with 
$\rm T_{eff} = 8750$ K, log g $=$ 2.0, [M/H] $=$ -1.5 and $\rm V_t = 2.0$ km $s^{-1}$
which fit the profile best. Open diamonds and asterixs represent the model 
 $\rm T_{eff} = 8500$ K, log g $=$ 2.5, [M/H] $=$ -1.5 and 
$\rm V_t = 2.5$ km $s^{-1}$, which does not fit the observed profile. 
No noticable differences were observed in the theoretical 
$\rm H_{\beta}$ profiles by considering ANOVER 
models for the same atmospheric parameters and are shown by dashes for the first set of
parameters and asterix in the case of second set of parameters.}

\end{figure*}

\section{Results}

The mean abundances of ZNG 4
relative to the Sun (Anders and Grevesse 1989) are given in Table 2, 
together with the number of lines used in the analysis and the standard 
deviation of abundances estimated from individual species.

\setcounter{table}{1}
\begin{table*} [ht]
\begin{center}
\caption{Chemical composition of ZNG 4 in M13}
\begin{tabular}{|c|c|c|c|c|c|c|}
\hline

Element& no of lines 
& \multicolumn{2}{c|}{$\rm T_{eff}=8500$ K, log g $=$ 2.5}
&\multicolumn{2}{c|}{$\rm T_{eff}=8500$ K, log g $=$ 2.5}
& $\rm T_{eff}=8750$ K, log g $=$ 2.0 \\


&
& \multicolumn{2}{c|}{ $\rm V_t = 2$ km $\rm s^{-1}$ and }
& \multicolumn{2}{c|}{ $\rm V_t = 2$ km $\rm s^{-1}$ and }
& { $\rm V_t = 2$ km $\rm s^{-1}$ and } \\

& & \multicolumn{2}{c|} {$\rm [Fe/H] = -1.5$ }
& \multicolumn{2}{c|}{$\rm [Fe/H] = -1.5$ }
& {$\rm [Fe/H] = -1.5$ } \\

&
& \multicolumn{2} {c|} {MOOG}
& \multicolumn{2} {c|} {WIDTH}
&  {MOOG} \\

\cline{3-7}

&  & $ \rm [X/H] \pm \sigma $ & [Ele/Fe] & $ \rm [X/H] \pm \sigma $ &
[ Ele/Fe] & $\rm [X/H] \pm \sigma $    \\

\hline
\hline

He I & 1 & -0.55 $\pm$ 0.20 &  & -0.68 $\pm$ 0.20 &      
& -1.13 $\pm$ 0.31   \\
Na I & 2 & -1.28 $\pm$ 0.10 & +0.21 & -1.30 $\pm$ 0.10  &  +0.20  
& -0.79 $\pm$ 0.11   \\
Mg I & 3 & -1.46 $\pm$ 0.07 & +0.03 & -1.48 $\pm$ 0.05 &  +0.02  
& -1.03 $\pm$ 0.16  \\
Mg II& 2 & -1.56 $\pm$ 0.14 & -0.07 & -1.59 $\pm$ 0.09 &  -0.09  
& -1.83 $\pm$ 0.15  \\
Si II& 4 & -1.10 $\pm$ 0.15 & +0.39 & -1.20 $\pm$ 0.12 &  +0.30  
& -1.31 $\pm$ 0.24  \\
Ca I & 2 & -0.80 $\pm$ 0.35 & +0.69 & -0.93 $\pm$ 0.25 &  +0.57  
& -0.09 $\pm$ 0.33  \\
Ca II& 1 & -0.49 $\pm$ 0.10 & +1.00 & -0.62 $\pm$ 0.10 &  +0.88  
& -0.32$\pm$ 0.10  \\
Sc II& 7 & -0.99 $\pm$ 0.25 & +0.50 & -1.04 $\pm$ 0.25 &  +0.45  
& -0.73$\pm$ 0.25  \\
Ti II& 51& -0.75 $\pm$ 0.18 & +0.74 & -0.83 $\pm$ 0.25 &  +0.67  
& -0.59 $\pm$ 0.19  \\
Cr II& 8 & -1.41 $\pm$ 0.12 & +0.08 & -1.47 $\pm$ 0.12 &  +0.03  
& -1.38$\pm$ 0.12  \\
Fe I & 13& -1.48 $\pm$ 0.15 & +0.02 & -1.53 $\pm$ 0.26 &  +0.02  
& -0.99$\pm$ 0.14  \\
Fe II& 28& -1.50 $\pm$ 0.21 & +0.00 & -1.46 $\pm$ 0.22 &  +0.02  
& -1.52 $\pm$ 0.22  \\
Sr II& 1 & -2.19 $\pm$ 0.10 & -0.70 & -2.21 $\pm$ 0.10 &  -0.71  
& -1.65 $\pm$ 0.12  \\
Y II & 2 & -1.20 $\pm$ 0.17 & +0.29 & -1.30 $\pm$ 0.12 &  +0.37  
& -0.64$\pm$ 0.14  \\
Ba II & 2 & -1.21 $\pm$ 0.05 & +0.28 & -1.23 $\pm$ 0.04 &  +0.26  
& -0.61$\pm$ 0.10  \\

\hline
\end{tabular}
\end{center}
\end{table*}

\setcounter{table}{2}
\begin{table*} [ht]
\begin{center}
\caption{Comparison of the abundances of ZNG4 with the abundances of M13
BHB and RGB stars}
\begin{tabular}{|c|c|c|c|c|c|}
\hline
Element&ZNG4&M13/ J11  (BHB) $^{*}$& M13/WF4-3485(BHB) $^{*}$ &
M13/ L262  (RGB) $^{*}$ & Mean \\
  & [Fe/H]$=$ -1.49 & [Fe/H]$=$ -1.82 & [Fe/H]$=$ $+$ 0.02 &
 [Fe/H]$=$ -1.61 & RGB abundances $^{*}$ \\
  &  $\rm T_{eff}=8500$ K & $\rm T_{eff}= 7681$ K & $\rm T_{eff}= 12750 $ K 
& $\rm T_{eff}= 4160$ K  & in M13 \\
 &  log g $=$ 2.5 & log g $=$ 3.1 & log g $=$ 4.1 & log g $=$ 0.50 & \\
\hline
\hline
   & [X/H]&[X/H]&[X/H]&[X/H]&[X/H]\\
He & $-$0.55 & $<$ $-$0.26 & $-$ 1.49 $\pm$ 0.17  
&  ...& ... \\
Na & $-$1.28 & ...               &  ...  
& $-$1.27 $\pm$ 0.11 & $-$1.37 $\pm$ 0.04 \\
Mg & $-$1.51 & $-$1.50 $\pm$ 0.16  & $-$ 1.62 $\pm$ 0.14 
& $-$1.51 $\pm$ 0.14 & $-$1.46 $\pm$ 0.03  \\
Si & $-$1.10 &  $<$ $-$1.23        & $-$ 1.43 $\pm$ 0.07 
& $-$1.16 $\pm$ 0.14 & $-$1.30 $\pm$ 0.02 \\
Ca & $-$0.64 &  $-$1.72 $\pm$ 0.11 & $-$ 1.66 $\pm$ 0.15 
& $-$1.39 $\pm$ 0.14 & $-$1.34 $\pm$ 0.01  \\
Sc & $-$0.99 &  $-$1.63 $\pm$ 0.10 & $<+$ 1.10  
&  ...   & $-$1.67 $\pm$ 0.01   \\
Ti & $-$0.75 &  $-$1.32 $\pm$ 0.06 & $-$ 0.53 $\pm$ 0.16 
& $-$1.30 $\pm$ 0.20 & $-$1.32 $\pm$ 0.02 \\
Cr & $-$1.41 &  $-$1.71 $\pm$ 0.12 & $<-$ 0.12  
& ...  &   ...           \\
Fe & $-$1.49 &  $-$1.82 $\pm$ 0.10 & $+$ 0.02 $\pm$ 0.20 
&  $-$1.61 $\pm$ 0.10 &  $-$1.60 $\pm$ 0.01 \\
Sr & $-$2.19 &  $-$2.03 $\pm$ 0.10 & $<$ + 1.75  
& ... &  ...   \\
Y  & $-$1.20 &   $<$ $-$0.97       & $<$ + 2.92  
& ... & $-$ 1.61 $\pm$ 0.04  \\
Ba & $-$1.21 &  $-$1.68 $\pm$ 0.10 & $<$ + 3.26  
& ...  & $-$ 1.53 $\pm$ 0.05  \\
\hline

\end{tabular}
\end{center}
\vspace{0.2cm}
\thanks{ *  Abundances of M13 BHB stars are from Behr (2000c) and
abundances of M13 RGB star are from Cavallo \& Nagar (2000). In the last
column, the mean abundances of elements from Na to Fe in M13 RGB stars 
are from Kraft et al.(1997) and that of Y and Ba are from Armosky et al.
(1994). } \\
\end{table*}

\begin{figure*}
\centering
\includegraphics[width=8.5cm,angle=90]{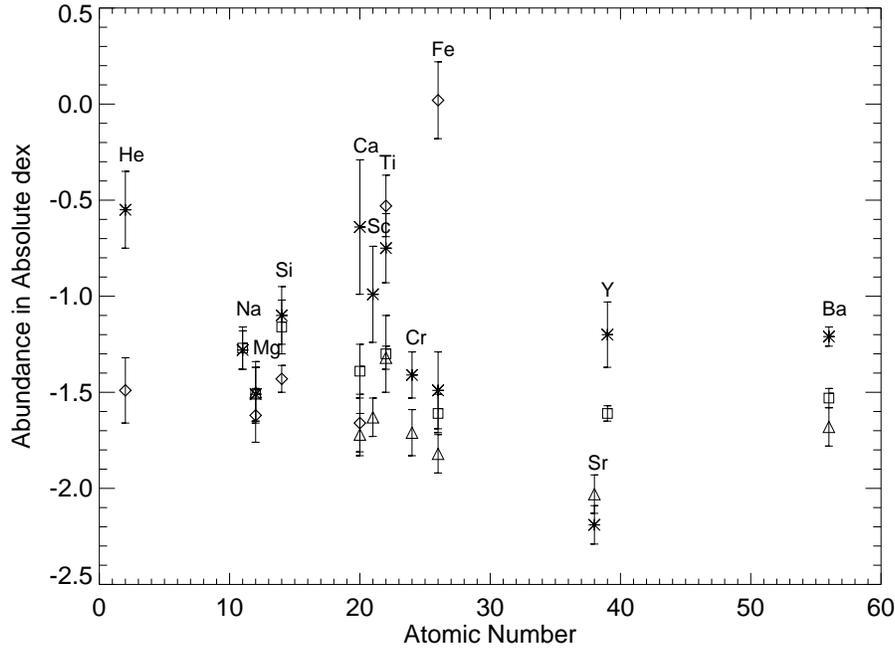}
\caption{
The abundances of different elements in ZNG 4 
(represented by asterix) compared with M13/ J11 ( BHB star; triangles), 
M13/WF4-3485 (hot BHB star; diamonds) and that of M13/ L262  
(RGB star; squares) with their error bars (Table 3). In the cases
where only upper limit of the abundances are given (3rd and 4th column of
Table 3), those elements are not shown in the above figure. 
Average abundances of Y and Ba in M13 RGB stars are from Armosky et al. 
(1994).
}
\end{figure*}

Analysis of Mg lines gives $\rm [Mg/H] = -1.5 $ which is the same as the M13
cluster metallicity. (The equivalent widths of the Mg II lines at 4481 $\AA$ 
(Table 1)
were obtained by a multiple gaussian fit to the lines in the observed 
spectrum).
Silicon is overabundant compared to iron 
($\rm [Si/Fe]=+0.4$). Calcium and titanium are found to be overabundant 
($\rm [Ca/Fe]=+0.8$  and  $\rm [Ti/Fe]=+0.75$).  There is a 0.5 dex 
difference in the abundances derived from the Ca I lines at 4226.73 $\AA$ and 
4454.78 $\AA$ (Table 1). However, the abundance of Ca derived from the Ca II line
at 5019.97 $\AA$ is in agreement with that derived from the Ca I 4454.78 $\AA$ 
line.
The reason for the deviation in the abundance derived from the Ca I 4226.73 $\AA$
line is not clear. It may be due to the relatively low signal to noise ratio
of the data around this wavelength range.
 There seems to be no interstellar contribution to the Ca II line.
Since the star has a radial velocity of -257 km $s^{-1}$, the stellar 
lines are well 
separated from the lines of interstellar origin. 
The abundance of Cr and Fe ($\rm [Fe/H] = -1.48$ and $\rm [Cr/Fe]=+0.09$) 
are found to be close to the metallicity of the cluster. 
On the other hand, Sc is found to be overabundant ($\rm [Sc/Fe]=0.51$).
Na lines show an overabundance of 0.2 dex. 
We have detected one line of Sr II, two lines of Y II and two lines of Ba II.
Sr seems to be underabundant ($\rm [Sr/Fe]=-0.70$), while Y and Ba are 
overabundant ($\rm [Y/Fe]=+0.29$ and $\rm [Ba/Fe]=+0.28$).

We have detected the He I line at 4471.47 $\AA$, which yields an abundance
of $\rm \log \epsilon (He) = 10.44$  which implies an underabundance of 0.55 dex compared to the
solar value. This is in agreement with the underabundane of He found in hot
BHB stars (Moehler, 1999, Moehler et al. 2003 ).

We have not detected C, N and O lines in our spectrum of ZNG 4. 
Assuming an equivalent width of 5 m$\AA$ as the detectable limit in our
spectrum of ZNG 4, we find the upper limit of $\rm [C/Fe]$ to be +0.32 dex
(based on the C I 5052.17 $\AA$ line ), that of  $\rm [N/Fe]$ to be +1.15 dex 
( based on the N I 4214.80 $\AA$ line ) and that of $\rm [O/Fe]$ to be 
+0.01 dex (based on the O I 6156.78 $\AA$ line ).
Globular cluster stars show anticorrelation 
of sodium and oxygen abundances (Kraft et al. 1997). In ZNG 4 we find 
enhancement of sodium, therefore we expect an underabundance of oxygen.
Also star to star abundance variations in the 
light elements C, N, O, Na, Mg and Al occur among the bright giants of a
number of globular clusters (Ivans et al. 1999). The absence of C, N and O
lines in our spectrum of ZNG 4 may be due to the underabundance of these
elements. A much higher resolution and high signal to noise ratio spectrum of
ZNG 4 may reveal the lines of C, N and O lines if present.

\section{Discussion and Conclusions}

     The chemical composition of ZNG 4 shows significant deviations
in element abundances from the expected metallicity of M13. 
Ti, Ca, Sc and Ba are found to be relatively overabundant (Figure 5)
compared to RGB stars and cool BHB stars of M13,
whereas the abundances of Mg, Cr and Fe are in agreement with the cluster 
metallicity. 

From a study of 22 M13 G-K giants, Kraft et al. (1993, 1997) found the 
abundances of Fe, Sc, V and Ni to be [Fe/H] = -1.49, 
[Sc/Fe] = -0.10, [V/Fe] = 0.00 and [Ni/Fe] = -0.04. They found Ca and Ti
to be mildly overabundant ([Ca/Fe] = +0.24 and [Ti/Fe]= +0.29). Si was 
overabundant by +0.34 dex. Study of M13 giants by Armosky 
et al. (1994) yields the average abundance of Fe, Y and Ba : 
[Fe/H] = - 1.49 , [Y/Fe] = -0.12 and [Ba/Fe] = -0.04, 
whereas in ZNG 4 we find significant
overabundance of Ca, Sc, Ti, Y and Ba compared to that found in 
M13 giants. Also, overabundance of Na [+0.2 dex] and absence of O lines 
support the anticorrelation of Na and O abundances found in M13 giants
(Kraft et al. 1997).

Behr et al.(1999) have studied the BHB stars in M13 on either side of the 
HB gap.
They find the photospheric compositions and stellar rotation rates to
vary strongly as a function of $\rm T_{eff}$ of the stars. 
Among the cooler stars in their sample, at $\rm T_{eff}$ of 8500K, the metal 
abundances are in rough agreement with the canonical cluster metallicity 
and the hotter stars with $\rm T_{eff}$ greater than 10000K show a deficiency
of He and enhancement of Fe, Ti, Cr by a factor of 300. However,
Mg remains at the canonical cluster metallicity. In ZNG 4 also, Mg abundance is 
in agreement with the M13 metallicity. Abundances similar to that found in 
BHB stars of M13 were also found in BHB stars
of M15 (Behr et al. 2000b) and BHB stars of NGC 6752 (Glaspy et al. 1989, 
Moehler et al.1999). The abundance anomalies in these
BHB stars are most likely due to diffusion - the gravitational settling
of helium (Greestein et al. 1967) and radiative levitation of metal
atoms (Michaud et al. 1983). Rotational velocities ({\it v sini}) appear to
have a bimodal distribution in cooler BHB stars, whereas the hotter
BHB stars with $\rm T_{eff}$ greater than 10000K are found to be slow 
rotators.

However, the abundances and rotational velocity of ZNG 4 are contrary to what 
is expected for stars on the red side of the HB gap at 11,000 K. It is a slow 
rotator ($v sin i =  7$ km ${sec}^{-1}$).  It shows underabundance of He and 
overabundances of Ti, Ca, Sc and Ba.
In Table 3 and Figure 5, we have compared the 
abundances of ZNG 4 with the abundances of M13 BHB stars of 
$\rm T_{eff}$ 7681 K (J 11)
and $\rm T_{eff}$ 12,750 K  (WF4-3485) (Behr, 2000c) and with the 
abundances of the RGB star L262 (Cavallo and Nagar, 2000). In Table 3, the last
column shows the mean RGB abundances in M13, where elements from Na to Fe
are taken from Kraft et al.(1997) and Y and Ba abundances are from 
Armosky et al. (1994).
It is evident from the abundances 
listed in Table 3 and from Figure 5 that ZNG 4 shows 
overabundance of metals 
compared to that of a M13 RGB star and also when compared to that of a M13 BHB 
star of similar temperature.
These results indicate that in ZNG4, diffusion and radiative levitation 
of elements may be in operation.
Slowly 
rotating HB stars are also seen on the cooler side of HB gap, but abundance 
anomalies start from 11,000 K (Moehler et al. 1999, Behr et al. 2000b). 
This implies that ZNG 4 may have the properties of the stars on
the blue side of the HB gap although it has $\rm T_{eff}$ of 8500 K.
In this regard, more accurate determination of abundances of these elements
in ZNG 4 and similar stars in M13 is needed to confirm our results and
conclusions.

This may be explained in two ways. One is that, for some stars in M13, 
the onset of 
diffusion seems to start at lower $\rm T_{eff}$ ($\approx $ 8500 K). 
The
other argument would be that the star has evolved from the blue side of 
the HB 
gap and is moving towards the red with higher luminosity as indicated by the
post-HB evolutionary tracks of Gingold (1976) and Dorman et al.(1993).

 The BHB stars hotter than 11500K typically show strong
photospheric helium depletions due to gravitational settling
(Moehler et al. 2000 \& 2003). The calculations of Michaud
et al. (1983) indicate that helium depletion should be
accompanied by photospheric enhancement of metals, since the
same stable atmosphere that permits gravitational settling
also permits the levitation of elements with large radiative
cross sections. The depletion of helium and overabudance of
some of the metals in the photosphere of ZNG 4 is in qualitative agreement
with the calculations of Michaud et al. (1983).

 Recently,
Turcotte et al. (1998) and Richer et al. (2000) made diffusion
simulations to
explain the abundance patterns of chemically peculiar A and F stars.
Their predicted abundance patterns are qualitatively similar to
that found in ZNG 4. However, none of the recent diffusion studies
treated the cases of BHB stars and post-HB stars. This phenomenon
may be related to the disappearance of surface convection and
hence to the formation of a stable stellar atmospheres. HB stars
 and post-HB stars  cooler than $\rm T_{eff}$ $=$ 6300K have deep convective
envelopes (Sweigart 2002).  Hotter than this temperature the envelope
convection breaks into distinct shells associated with the
ionization of H and He. Note that the surface convection
disappears at 11000K ( Sweigart 2002) and BHB stars hotter than
this show moderate to severe abundance anomalies. 

ZNG 4 has a V magnitude of 13.964 (Paltrinieri et al. 1998). Considering 
the distance modulus of M13 to be 14.42 and $\rm E(B-V)$ towards M13 
to be 0.02 (Kraft and Ivans 2003), we estimated the absolute magnitude ($\rm M_v$)
of the star to be $ -0.522$. For stars with $\rm T_{eff}$ around 8500 K, 
the bolometric correction (BC) is negligible (Flower 1996). 
Considering $\rm BC=0$, we get the bolometric magnitude ($\rm M_{bol}$) to 
be $-0.522$, 
which corresponds to a luminosity of 127 $\rm L_{\odot}$
[ $\rm log {\frac {L} {L_{\odot}}} = 2.18$].
Using the equation connecting the mass, effective temperature and bolometric
magnitude, we find the surface gravity, log g = 2.6 (assuming the mass of ZNG 4
to be 0.5 $\rm M_{\odot}$), which agrees well with the value estimated from
the analysis of the spectrum of ZNG 4. 

The post-AGB star Barnard 29 [ $\rm log {\frac {L} {L_{\odot}}} = 3.3$]
which is a member of M13 is more luminous than ZNG 4.
The abundance pattern of ZNG 4 is very different from
that of the post-AGB star Barnard 29 (Conlon et al. 1994, Moehler et al. 
1998). The M13 BHB stars with $\rm T_{eff}$ around 8500 K have a luminosity 
of about 40 $\rm L_{\odot}$, whereas ZNG 4 is more luminous by about a 
factor of 3, which indicates that ZNG 4 can be classified as a supra 
horizontal branch star (post-HB).
Stars that lie 1.5 magnitude above the HB stars are classified 
as supra horizontal branch (SHB) stars in the photometric studies of M13 
(Zinn 1974) and NGC 6522 (Shara et al. 1998).
No detailed abundance analysis of SHB stars in globular clusters 
is available to compare with the abundances of ZNG 4. 

Since ZNG 4 is a
post-HB star and it has evolved from a hot BHB star stage and may
had severe abundance anomalies similar to those found in
the hot BHB stars of
M13 (Table 3). The present $\rm T_{eff}$ $=$ 8500K of ZNG 4 indicates that
thin layers of subsurface convection if present may have diluted the severe
abundance anomalies due to diffusion and radiative levitation that
took place during its hot BHB stage of evolution.  
 It is important
to derive the chemical composition of a significant sample of
post-HB stars hotter than 11000K and much cooler than 11000K
to further understand the role of diffusion, radiative
levitation, rotation and convection during the post-HB stage of evolution.

\begin{acknowledgements}
We would like to thank Dr. B.Dorman and Dr.B.Behr for helpful 
information on M13 BHB stars, Dr. F.R.Ferraro for kindly providing the table 
of B and V magnitudes of M13 cluster stars and the referee Dr.S.Moehler for 
helpful comments.

\end{acknowledgements}


\enddocument